\documentclass[a4paper,11pt]{article}
\usepackage{pos}
\usepackage{microtype}

\renewcommand{\vec}{\mathbf}
\newcommand{\op}{\mathcal{O}}

\title{Hadron spectroscopy and interactions}

\author*{Jeremy R.\ Green}

\affiliation{John von Neumann-Institut für Computing NIC,
  Deutsches Elektronen-Synchrotron DESY,\\
  Platanenallee 6, 15738 Zeuthen, Germany}

\emailAdd{jeremy.green@desy.de}

\abstract{In recent years, lattice QCD calculations of hadron
  spectroscopy have concentrated on resonances and shallow bound
  states detected via poles in two- and three-hadron scattering
  amplitudes. Hadron interactions have therefore become a key
  focus. In these proceedings, I review the current state of the art
  and recent advances in methods for studying hadron interactions via
  finite-volume spectroscopy and finite-volume quantization
  conditions. I will also review recent spectroscopy studies and
  results presented at Lattice 2025, with a focus on charmed mesons,
  the doubly charmed tetraquark, and the doubly bottom tetraquark.}

\FullConference{The 42nd International Symposium on Lattice Field Theory (LATTICE2025)\\
2-8 November 2025\\
Tata Institute of Fundamental Research, Mumbai, India\\}

\renewcommand{\logo}{\relax}

\renewcommand{\hookAfterAbstract}{%
  \par\bigskip
  \textsc{ArXiv ePrint}: \href{https://arxiv.org/abs/2602.23244}{2602.23244}
  \hfill DESY-26-031
}

\begin{document}
\maketitle

\makeatletter
\g@addto@macro\bfseries{\boldmath}
\makeatother

\section{Introduction}

\begin{figure}[b]
  \centering
  \includegraphics[scale=0.9]{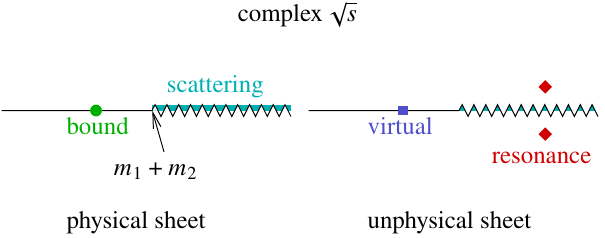} \hfill
  \includegraphics[scale=0.9]{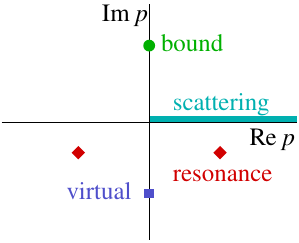}
  \caption{Locations of physical scattering and possible poles for a
    two-particle partial wave amplitude. Left: on the complex
    centre-of-mass energy plane, which has physical and unphysical
    Riemann sheets resulting from a branch cut starting at
    threshold. Right: on the complex scattering momentum plane, which
    opens the cut via $\sqrt{s}=\sqrt{m_1^2+p^2}+\sqrt{m_2^2+p^2}$.}
  \label{fig:poles}
\end{figure}

The goal of hadron spectroscopy is understanding the composite states
formed from quarks and gluons. In the last decade, the focus of these
studies on the lattice has shifted from hadrons that are stable in QCD
to those that are unstable, decaying to other hadrons. Formally, both
stable and unstable states exist as poles in the partial wave
scattering amplitudes of the stable hadrons to which they couple; see
Fig.~\ref{fig:poles}. It is thus necessary to study scattering states
in order to investigate resonance and virtual-state poles.

Experimental studies of the hadron spectrum are still very active, in
part because of the many exotic hadrons that have been discovered. In
the broadest sense, these are states not predicted by the standard
quark model in which mesons are quark-antiquark states and baryons are
three-quark states. Beyond simple disagreements with the precise quark
model predictions, this also includes states that contain four or more
quarks and mesons with $J^{PC}$ not allowed by the quark model.

In addition to the considerable interest in exotic hadrons, lattice
QCD calculations are also being used to study conventional unstable
hadrons, including the lightest ones such as $f_0(500)$, $\rho(770)$,
and $\Delta(1232)$. Multihadron physics is interesting in its own
right: for instance, two or three pions can appear as intermediate
states in the hadronic vacuum polarization or as final states in kaon
decay. Finally, the techniques of hadron spectroscopy can be used to
better isolate states for hadron structure studies, as discussed in
the plenary talk by Barca~\cite{Barca:2026juc}.

This review is divided into two parts. Section~\ref{sec:methods}
reviews methods for hadron spectroscopy and recent developments. Some
specific hadron systems that have been investigated in the literature
and at this conference are discussed in Section~\ref{sec:hadrons}. An
outlook is given in Section~\ref{sec:outlook}.

\section{Methods}
\label{sec:methods}

The standard approach for studying scattering amplitudes (and stable
hadrons) is finite-volume spectroscopy. This can be organized into
four main steps:
\begin{enumerate}
\item Computing matrices of two-point correlation functions,
  \begin{equation}
    C_{ij}(t) \equiv \left\langle
      \op_i^{\vphantom{\dagger}}(t) \op_j^\dagger(0) \right\rangle,
      \qquad 1 \leq i,j \leq N_\text{op},
  \end{equation}
  with the $N_\text{op}$ interpolating operators $\op_i$ chosen to
  have definite flavour, momentum, and rotational quantum numbers.
\item Estimating the low-lying finite-volume energies $E_n$ from the
  time dependence of $C_{ij}(t)$.
\item Applying finite-volume quantization conditions to constrain
  models of partial wave scattering amplitudes using the finite-volume
  energies.
\item Analytically continuing the amplitudes to find poles
  corresponding to hadrons (bound states, virtual states, or
  resonances).
\end{enumerate}
For each of these steps, the state of the art and recent developments
will be discussed in a subsection below.

An alternative approach, used less widely, is the HAL QCD method. This
requires an analogous series of steps:
\begin{enumerate}
\item Computing hadron-hadron correlation functions,
  \begin{equation}
    F(\vec r,t) \equiv \int d^3\vec x \left\langle
      \op_{H_1}(\vec x,t) \op_{H_2}(\vec x+\vec r,t) \op_{H_1H_2}^\dagger(0)
      \right\rangle,
  \end{equation}
  where $\op_{H_i}$ are local interpolating operators for the two
  interacting hadrons separated by $\vec r$ and $\op_{H_1H_2}$ is an
  interpolating operator for the two-hadron system.
\item Estimating the nonlocal interaction potential
  $U(\vec r,\vec r')$ from the correlation function. This has a scheme
  dependence coming from the choice of local interpolating operators
  and is usually approximated using the leading-order term in a
  derivative expansion,
  $U(\vec r,\vec r') \approx V^\text{LO}(r)\delta^{(3)}(\vec r-\vec
  r')$.
\item Fitting the potential using model functions $V(r)$.
\item Solving the Schrödinger equation using the potential models to
  determine bound state energies and partial wave amplitudes, which
  are scheme independent.
\end{enumerate}
As this method is likely to have quite different systematic
uncertainties, it provides a valuable alternative to finite-volume
spectroscopy. However, the focus of this section is on the standard
approach.

\subsection{Correlation functions}

The simplest and oldest methods for computing correlation functions of
interpolating operators built from quark fields are to use (smeared)
point sources or Coulomb-gauge wall sources. These can be effective
for determining the masses of isolated stable hadrons; however, the
restricted spatial structure of the interpolating operators makes a
calculation of the full spectrum of scattering states
infeasible. These methods are also sometimes associated with
asymmetric correlation functions, i.e.\ with differing source and sink
interpolating operators, which can suffer from non-positive-definite
correlation functions and non-monotonic behaviour of effective
energies. The latter can risk misidentifying even ground-state
energies.

One can produce a greater variety of interpolating operators,
including nonlocal ones that couple well to scattering states, by
combining point sources with sequential- and stochastic-source methods
such as in Refs.~\cite{Alexandrou:2017mpi, Silvi:2021uya,
  Barca:2022uhi}.

All-to-all methods allow for more reuse of data and a better cost
scaling with the number of interpolating operators. In recent years,
position-space sparsening methods have been
explored~\cite{Detmold:2019fbk, Li:2020hbj}. These are based on
replacing the spatial sums used for momentum projection with reduced
sums over regular grids or random subsets of the lattice. If the
subsets are sufficiently small, it is feasible to compute and save the
all-to-all (smeared) quark propagator between points belonging to the
subsets. Since smearing has a physically large footprint, a
substantial sparsening is possible with negligible loss of signal
quality.

The most widely used all-to-all method is
\emph{distillation}~\cite{HadronSpectrum:2009krc, Morningstar:2011ka},
which is based on computing the all-to-all propagator within a
subspace of dimension $4N_\text{LapH}$ on each timeslice. This
``distillation subspace'' is formed from the lowest $N_\text{LapH}$
eigenmodes of the three-dimensional Laplacian, and the corresponding
Laplacian-Heaviside (LapH) smearing of quark fields is a projection
onto this subspace. In recent years, some variants and improvements
have been explored. One avenue is improving the smearing using
\emph{quark distillation profiles}, which allow for a more optimal
tuning within the distillation space~\cite{Knechtli:2022bji}; an
update combining this with covariant derivatives was presented by
Urrea-Niño~\cite{Urrea-Nino:2026cjj, UrreaNino_Lat25}. Going in a
different direction, the distillation subspace can be varied: one
option is to use plane waves in Coulomb
gauge~\cite{Sugiura:2022bgo}. The latter was used in a study of the
nuclear spin-orbit force within the HAL QCD method, presented by
Sugiura~\cite{Sugiura_Lat25}.

One of the drawbacks of distillation is the high cost of contractions
for local interpolating operators with four or more quarks at the same
point, which scales as $N_\text{LapH}^5$ or worse. Two solutions have
been developed: the first constructs a localized basis for the
distillation space~\cite{Lang:2024syy} and the second incorporates
position-space sparsening~\cite{Stump:2025owq}. Stump presented an
application of the latter to the doubly charmed
tetraquark~\cite{Stump:2026xvg}.

A universal problem for multihadron correlation functions is the
possibly large number of Wick contractions. Naïvely, the cost is
factorial in the number of quarks of each flavour; however, many of
the intermediate expressions appear repeatedly. Much work has been
done to develop algorithms that reduce this cost~\cite{Doi:2012xd,
  Detmold:2012eu, Horz:2019rrn, Chen:2023zyy, Chakraborty:2024oym,
  Selvitopi:2025mlx}; one method was presented by
Chakraborty~\cite{Chakraborty_Lat25}.

For finite-volume multihadron spectroscopy, scaling to large volumes
is also very challenging. In distillation, $N_\text{LapH}$ must be
scaled proportionally to the physical three-dimensional volume $L^3$
to maintain a constant smearing radius. Since multi-meson systems
involve only rank-2 tensors in the distillation space, their
contraction cost scales as $N_\text{LapH}^3\propto L^9$. Systems with
one or two baryons have a cost scaling with
$N_\text{LapH}^4\propto L^{12}$~\cite{Horz:2019rrn,
  Green:2021qol}. The density of states is also a problem, since
$N_\text{op}$ should be scaled with the number of states in the energy
interval of interest and the naïve contraction cost is proportional to
$N_\text{op}^2$. For fixed total momentum and a fixed energy interval,
the number of two-hadron states asymptotically scales proportionally
to $L^3$, which increases to $L^6$ for three-hadron
states. Furthermore, as the gaps between energy levels become smaller,
we need higher precision to get useful constraints from finite-volume
quantization conditions. Altogether, in the large-volume regime, we
get an extremely rapid cost scaling with box size. New ideas may be
helpful here.

\subsection{Lattice spectrum}

The starting point for estimating energy levels from a matrix of
two-point correlation functions is its spectral decomposition:
\begin{equation}
  C_{ij}(t) = \sum_n e^{-E_n t} Z_{in}^{\vphantom{*}} Z_{jn}^*,\qquad
  Z_{in} \equiv \left\langle\Omega \middle| \op_i \middle| n \right\rangle,
\end{equation}
where thermal effects have been neglected. Having an
$N_\text{op}\times N_\text{op}$ matrix of correlation functions is
very important, since it is possible to identify up to $N_\text{op}$
near-degenerate energy levels via the rank of the matrix that
multiplies the exponential.

The standard analysis of a correlator matrix is to solve a generalized
eigenvalue problem (GEVP)~\cite{Michael:1985ne, Luscher:1990ck},
\begin{equation}
  C(t) v_n(t,t_0) = \lambda_n(t,t_0) C(t_0) v_n(t,t_0).
\end{equation}
This is a \emph{variational method} in that it implicitly finds an
optimal linear combination
$\overline \op_n = \sum_i (v_n^\dagger)_i \op_i$ for coupling to state
$n$, and $\lambda_n(t,t_0)\approx e^{-E_n(t-t_0)}$. It has been
proven~\cite{Blossier:2009kd} that for a particular choice of the
reference time $t_0$, one gets estimators for the low-lying energy
levels with good asymptotic behaviour:
\begin{equation}
  - \frac{d}{dt} \log \lambda_n(t,t_0) = E_n
  + O\left( e^{-(E_{N_\text{op}}-E_n)t} \right).
\end{equation}
By increasing $N_\text{op}$, one can obtain a faster convergence to
the energies.

\begin{figure}
  \centering
  \includegraphics{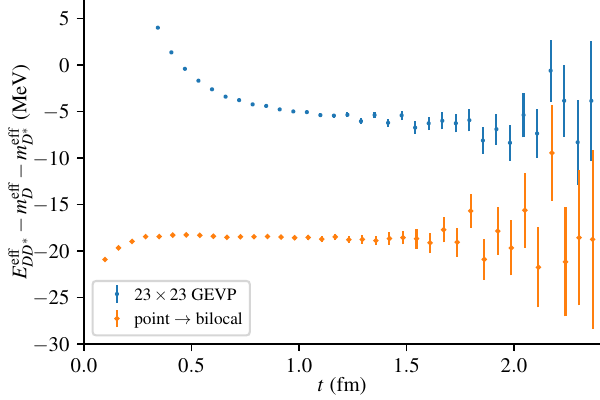}
  \caption{Difference between the ground-state effective energy of a
    $DD^*$ system and the sum of effective masses of a $D$ and a $D^*$
    meson. Blue circles are obtained from solving a GEVP for a
    symmetric correlator matrix with $N_\text{op}=23$; orange diamonds
    are obtained from a single asymmetric correlation function with
    local creation operator and bilocal annihilation operator.}
  \label{fig:Tcc_plateaus}
\end{figure}

Figure~\ref{fig:Tcc_plateaus} illustrates the importance of using a
large symmetric correlator matrix and the danger of using a single
asymmetric correlator. This particular example of the latter produces
a very long and stable plateau, lasting for more than 1~fm with error
below 1~MeV. The GEVP produces a completely different, much later and
worse looking, plateau. Naïvely looking at this plot, one is tempted
to prefer the asymmetric correlator. However, the relevant length
scale is the inverse of the energy gap $\Delta E$ governing
corrections to the effective energy. For this two-hadron system and a
single correlation function, $\Delta E^{-1}\approx 6$~fm, so that a
plateau of length 1~fm is inadequate. The GEVP, on the other hand,
increases the effective energy gap significantly, shrinking the
relevant length scale severalfold. For this reason, one should expect
the GEVP plateau to be more accurate, and the asymmetric plateau,
disagreeing with it, is probably incorrect. Similar issues have
plagued past studies of baryon-baryon systems~\cite{Iritani:2016jie,
  Green:2025rel}; see the plenary talk by Nicholson on this
topic~\cite{Nicholson_Lat25, BaryonScattering:2025ziz}.

In recent years, there have been new efforts to better use the time
dependence of correlation functions or matrices to extract information
about the spectrum. One strategy is to apply regularized Laplace
filters, which suppress correlations and alter the spectral
weights~\cite{Portelli:2025lop}. Other methods aim to generalize
effective masses and GEVPs to incorporate data from many or all
source-sink separations $t$ rather than just a pair of late
ones~\cite{Wagman:2024rid, Hackett:2024nbe, Ostmeyer:2024qgu,
  Chakraborty:2024exj, Abbott:2025yhm, Ostmeyer:2025igc,
  Tsuji:2025zdn}: these include the new application of Lanczos-type
algorithms to the transfer operator $e^{-aH}$ as well as a revival of
older ideas such as the Prony and generalized pencil-of-function
methods. In the absence of noise and in suitable limits, these methods
are equivalent and yield estimators that converge more rapidly to
energy levels. The main difficulties come from having an exact
representation of a correlator containing noise, which generally does
not have a positive-definite spectral representation. Within this
family, Ostmeyer presented a truncated Hankel correlator method, which
uses rank-$k$ approximations of a Hankel matrix containing the full
correlator data [for a scalar correlator, the matrix is
$H_{ij}(t)=C(t+(i+j)a)$] as closed-form alternatives to multi-state
fits~\cite{Ostmeyer:2025igc, Ostmeyer_Lat25}. With some time having
passed since the Lanczos proposal, it is now seeing some applications:
Perry presented a study of nucleon-nucleon data using this method
including two-sided bounds the method provides for
energies~\cite{Detmold:2026shh, Detmold:2026vjx, Perry_Lat25}.

\subsection{Finite-volume quantization}
\label{sec:quantization}

Quantization conditions provide the relationship between the
finite-volume energies that we can compute on the lattice and the
infinite-volume scattering amplitudes and bound states that are
studied in experiments. For single hadrons or deeply bound states, the
relationship is simple: corrections are exponentially suppressed and
the finite-volume state is a good approximation to the infinite-volume
state. For two-hadron systems, there is now a seemingly mature
formalism following years of developments: in addition to dealing with
arbitrary coupled channels and arbitrary spin, one can also compute
decays, transitions, and matrix elements. A toy-model investigation of
the latter for a range of binding energies was presented by
Ortega-Gama~\cite{Moscoso:2026wmz, OrtegaGama_Lat25}.

Three-particle quantization conditions are increasingly being applied
to real data (see Section~\ref{sec:three_mesons}), although they are
not yet fully general. Sharpe presented quantization conditions for an
$N\pi\pi$ system with maximal isospin~\cite{Hansen:2025oag,
  Sharpe_Lat25}; see also the plenary talk reviewing three-particle
amplitudes by Sharpe~\cite{Sharpe:2026mtt}. Going beyond three
particles, a perturbative study of a finite-volume four-pion system
was presented by Mukherjee~\cite{Mukherjee_Lat25}.

\begin{figure}
  \centering
  \includegraphics[width=0.8\textwidth]{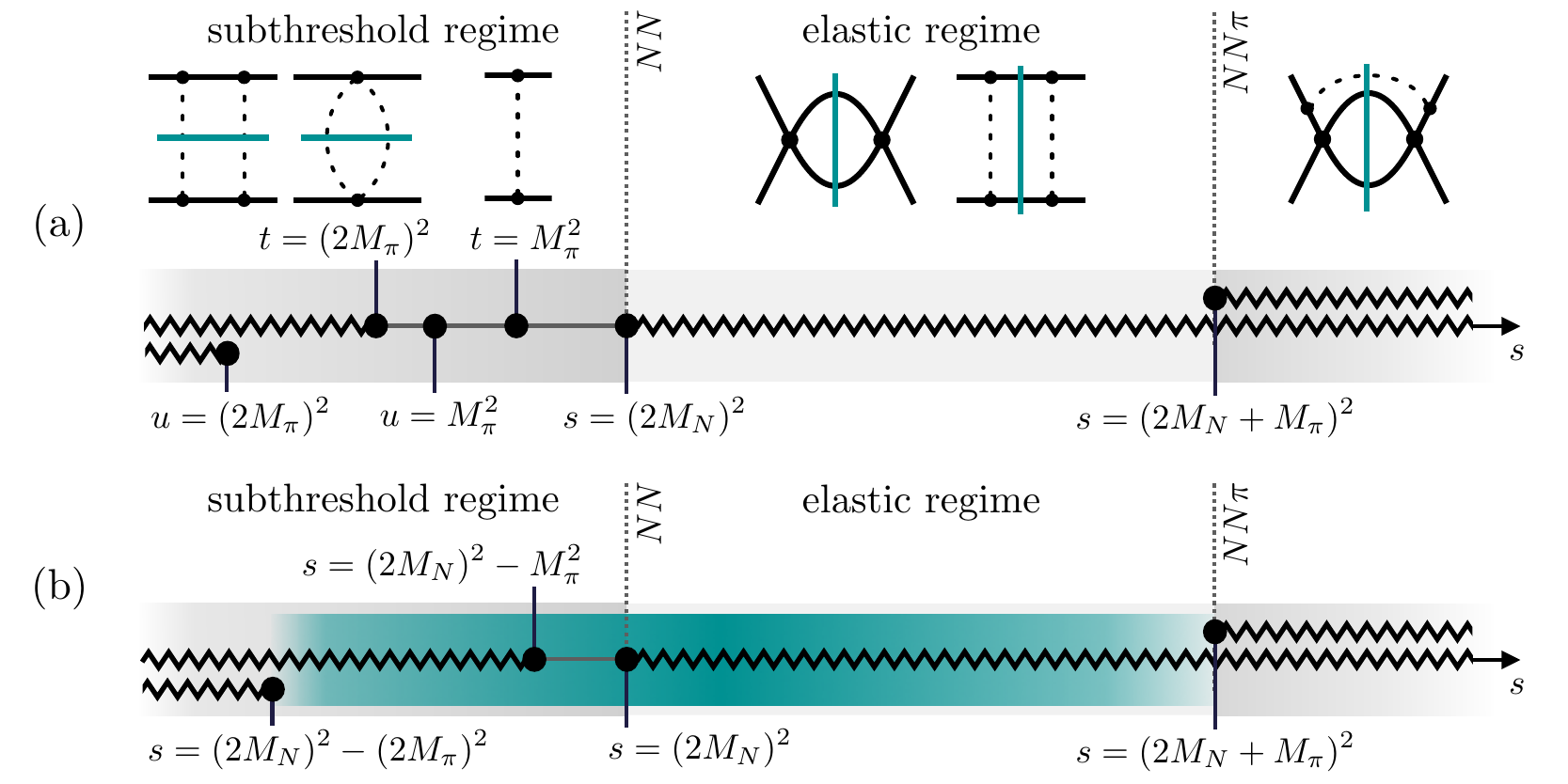}
  \caption{Analytic structure of the $NN$ scattering amplitude: (a)
    poles and cuts at fixed scattering angle and (b) cuts after
    partial-wave projection. Standard two-body quantization conditions
    only deal with the elastic two-particle branch cut. Reproduced
    from Ref.~\cite{Raposo:2023oru} under the Creative Commons
    Attribution License
    (\href{https://creativecommons.org/licenses/by/4.0/}{CC-BY-4.0}).}
  \label{fig:left_hand_cuts}
\end{figure}

Returning to the two-particle case, there have been recent
developments despite the maturity of the formalism. These have been
instigated by the realization that Lüscher's quantization conditions
break down below threshold in the presence of \emph{left-hand
  cuts}~\cite{Green:2021qol, Du:2023hlu}. Such cuts are shown in
Figure~\ref{fig:left_hand_cuts} and occur due to the exchange of
particles in crossed channels ($t$ and $u$ channel). They can lie
close to the threshold in cases when two heavy hadrons exchange a
light hadron; in particular, this was noted as a problem due to pion
exchange in two-baryon and $DD^*$ systems.

By now, five different solutions have been developed. In chronological
order:
\begin{enumerate}
\item Using the plane-wave basis to quantize a scattering equation
  such as a Lippmann-Schwinger equation in finite
  volume~\cite{Meng:2021uhz, Meng:2023bmz}. An application of this was
  presented by Shrimal~\cite{Shrimal:2025ues, Shrimal:2026bur}.
\item Modified Lüscher quantization conditions to obtain a
  short-distance quantity that is related to the scattering amplitude
  via integral equations~\cite{Raposo:2023oru, Raposo:2025dkb}.
\item Three-particle quantization, with the two-particle amplitude
  obtained via LSZ reduction~\cite{Hansen:2024ffk, Dawid:2024dgy}. An
  application of this was presented by Raposo~\cite{Alharazin:2026lno}.
\item Modified Lüscher quantization conditions applied to a modified
  effective range expansion~\cite{Bubna:2024izx, Bubna:2025gsd}.
\item Quantization conditions based on an $N/D$ representation of the
  amplitude~\cite{Dawid:2024oey}.
\end{enumerate}
As discussed in Section~\ref{sec:Tcc}, the three-particle approach is
particularly suited to systems such as $DD^*$, which becomes $DD\pi$
at the physical pion mass where $D^*$ is a resonance, since it works
smoothly across the transition from two to three particles.

\subsection{Amplitudes and analytic continuation}

A commonly-used strategy for describing scattering amplitudes is to
fit with a variety of simple analytic models, which can then be
trivially continued into the complex plane to find poles. Among the
models that yield good descriptions of the lattice spectra, the
scatter of results is used to estimate systematic uncertainty.

One way of improving on this approach is to impose stronger physics
constraints on the models. In Ref.~\cite{Rodas:2023nec}, crossing
symmetry of the $\pi\pi$ scattering amplitude (Roy equations) was used
to filter out unphysical models, yielding smaller systematic
uncertainty on the location of the $\sigma$-resonance pole.

An alternative strategy presented by Salg is to avoid explicit
analytic models~\cite{Salg:2025now, Salg_Lat25}. Instead, the
amplitudes are first constrained on the real axis using Bayesian
methods similar to Gaussian processes, then continued to the complex
plane using Nevanlinna-Pick interpolation.

\section{Hadron systems}
\label{sec:hadrons}

\subsection{Two light mesons}

For the simplest resonances involving pions and kaons, lattice
calculations are maturing. Refs.~\cite{Boyle:2024hvv, Boyle:2024grr}
presented a study of the $\rho(770)$ and $K^*(892)$ in $\pi\pi$ and
$\pi K$ scattering using one ensemble with physical quark masses; the
former was also studied at the physical pion mass by
Refs.~\cite{Fischer:2020yvw, Wang:2025hew}. We are thus approaching
the possibility of precision $\pi\pi$ physics, which calls for an
effort to control all sources of systematic uncertainty. Talks at this
conference went in different directions: Valois laid the groundwork
for $\pi\pi$ scattering with staggered fermions~\cite{Valois_Lat25},
Morandi studied $\pi\pi$ systems using spectral reconstruction
techniques valid above the inelastic threshold~\cite{Morandi:2026nll},
and Erben presented a study of the rare decay
$B\to K^*(\to K\pi)\ell^+\ell^-$~\cite{Erben:2026ylp}.

\subsection{Three light mesons}
\label{sec:three_mesons}

Lattice studies of three-meson systems have been rapidly advancing
beyond the first proof-of-concept calculations. Romero-López presented
a calculation of the maximal-isospin systems $\pi^+\pi^+\pi^+$,
$\pi^+\pi^+K^+$, $\pi^+K^+K^+$, and $K^+K^+K^+$ extending down to the
physical pion mass, with the case of pions showing agreement with
chiral perturbation theory~\cite{Dawid:2025zxc, Dawid:2025doq,
  RomeroLopez_Lat25}. Other collaborations have investigated three
pions with isospin zero~\cite{Yan:2024gwp}, one~\cite{Yan:2025mdm},
and two~\cite{Briceno:2025yuq, Feng:2026ixm}, which are challenging in
part because of $\rho$ and $\sigma$ resonances appearing in
two-particle subsystems. Mai presented the isospin-one
study~\cite{Mai_Lat25}, which identified the $J^{PC}=0^{-+}$
$\pi(1300)$ resonance at pion mass 305~MeV and extrapolated it to the
physical pion mass. For lighter pion masses, this would lie above the
$5m_\pi$ threshold where the three-particle quantization condition
breaks down. For a more detailed discussion of three-particle
scattering, see the review talk by Sharpe~\cite{Sharpe:2026mtt}.

\subsection{Charmed mesons}
\label{sec:charm_mesons}

\begin{figure}
  \centering
  \includegraphics{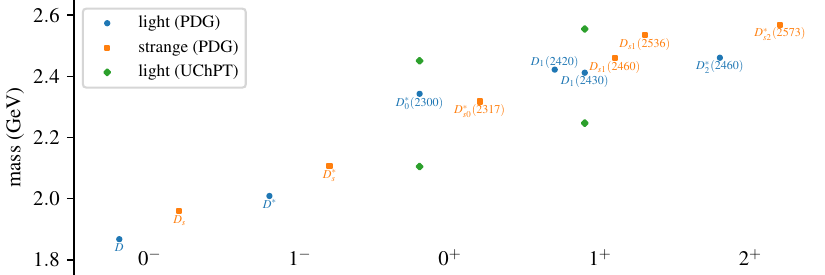}
  \caption{Low-lying $D$ and $D_s$ mesons for various $J^P$. Blue
    circles and orange squares indicate light and strange states
    listed in the PDG~\cite{ParticleDataGroup:2024cfk}; green diamonds
    show the two-pole structures in the UChPT analysis of
    Ref.~\cite{Du:2017zvv}.}
  \label{fig:D_mesons}
\end{figure}

Figure~\ref{fig:D_mesons} shows the experimentally observed low-lying
charm-antilight and charm-antistrange mesons. For $J^P=0^-$, $1^-$,
and $2^+$, the typical pattern can be seen: substituting a strange
instead of a light antiquark increases the mass by about 100~MeV. One
can also identify a pair of $1^+$ states that follow the same
pattern. However, two strange mesons have unusually low masses
compared with the corresponding light meson: $D_{s0}^*(2317)$ and
$D_{s1}(2460)$. According to studies in unitarized chiral perturbation
theory (UChPT), the resolution of this puzzle is that two of the light
mesons, $D_0^*(2300)$ and $D_1(2430)$, have been misidentified and
each of them is actually a pair of two distinct
poles~\cite{Du:2017zvv}. In the SU(3) flavour limit, scattering of a
$D_{(s)}$ meson and an octet meson belongs to the irreps
$\mathbf{\overline{3}}\otimes\mathbf{8} =
\mathbf{\overline{3}}\oplus\mathbf{6}\oplus\mathbf{\overline{15}}$;
one member of each of these pairs belongs to the conventional
$\mathbf{\overline{3}}$ representation and the other belongs to the
exotic $\mathbf{6}$ representation containing at least four
quarks. The latter also contains manifestly tetraquark states: an
$I=1$ $T_{c\bar s}$ (with some recent evidence for it at
LHCb~\cite{LHCb:2024iuo}) and an $I=0$ $T_{cs}$ with minimal quark
content $cs\bar u\bar d$. UChPT does not predict states in the third
representation, $\mathbf{\overline{15}}$, in contrast to
diquark-antidiquark models~\cite{Guo:2025vjw}. The ability of lattice
calculations to vary quark masses and study both SU(3)-symmetric
points and the breaking of SU(3) symmetry is thus a great asset for
disentangling these models and effective theories.

We first discuss $J^P=0^+$. In Ref.~\cite{Yeo:2024chk}, calculations
were done on three volumes at a heavy SU(3) point with
$m_\pi=m_K\approx 700$~MeV. In flavour $\mathbf{\overline{3}}$, they
found a bound state, in $\mathbf{6}$, they found a virtual state, and
in $\mathbf{\overline{15}}$, the behaviour was repulsive. This is
consistent with the UChPT prediction. Using previous
calculations~\cite{Moir:2016srx, Cheung:2020mql, Gayer:2021xzv}, one
can track some elements of these multiplets as the pion mass is
decreased to 391 and 239~MeV and SU(3) is broken. In the
$\mathbf{\overline{3}}$, the $D_0^*$ becomes a resonance, whereas the
$D_{s0}^*$ remains a bound state. (A bound $D_{s0}^*$ was previously
found in Refs.~\cite{Mohler:2013rwa, Lang:2014yfa, Bali:2017pdv,
  Alexandrou:2019tmk}.) In the $\mathbf{6}$, only the $T_{cs}$ was
studied and was found to remain a virtual state. Other recent studies
of $S$-wave $D\pi$ scattering are Ref.~\cite{Yan:2024yuq}, which
reaches down to the physical pion mass, and the poster presented by
Thoma~\cite{Thoma_Lat25}. Different models for tetraquark states can
also be probed via more significant departures from QCD. In
Ref.~\cite{Baeza-Ballesteros:2025iee}, simulations were done with
$N_c\in\{3,4,5,6\}$ and SU(4) flavour symmetry; virtual $T_{cs}$ and
$T_{c\bar s}$ states were only found for $N_c=3$, which disfavours a
compact tetraquark structure.

For $J^P=1^+$, Ref.~\cite{Lang:2025pjq} worked with broken SU(3)
symmetry and studied the coupled-channel system
$D^*\pi$--$D^*\eta$--$D^*_s\bar K$. Consistent with the UChPT picture,
three $1^+$ states were found: a bound state, a narrow resonance, and
a broad resonance. The only calculation with SU(3) symmetry was
presented by Gregory~\cite{Gregory:2025ium, Gregory_Lat25}. Although
it did not include a full scattering analysis, it found that for both
$J^P=0^+$ and $1^+$, the $\mathbf{6}$ was attractive and the
$\mathbf{\overline{15}}$ was repulsive, again favouring the molecular
picture of UChPT over a compact tetraquark structure.

\subsection{Doubly charmed tetraquark}
\label{sec:Tcc}

The $T_{cc}(3875)^+$ tetraquark has been of particular interest since
its discovery in the $D^0D^0\pi^+$ invariant mass spectrum at
LHCb~\cite{LHCb:2021vvq, LHCb:2021auc}. With minimal quark content
$cc\bar u\bar d$ and quantum numbers $I=0$, $J^P=1^+$, it lies
$360\pm 40\;^{+4}_{-0}$~keV below the $D^{*+}D^0$ threshold with a
width of $48\pm 2\;^{+0}_{-14}$~keV, making it the longest-lived
exotic hadron. For pion masses slightly above physical, the $D^*$
becomes stable and we can investigate the $T_{cc}$ as a possible
two-body $DD^*$ bound state in the $S$ wave.

An initial group of lattice calculations~\cite{Ikeda:2013vwa,
  Padmanath:2022cvl, Chen:2022vpo, Lyu:2023xro, Whyte:2024ihh}, done
at pion masses between 146 and 411~MeV for which the $D^*$ is stable,
found a virtual state that they identified as the $T_{cc}$. For
decreasing pion mass, the pole moved toward the $DD^*$ threshold where
it would become a bound state. In a coupled-channel analysis,
Ref.~\cite{Whyte:2024ihh} also found a resonance with strong coupling
to $D^*D^*$.

More recently, there have been more investigations of some sources of
systematic uncertainty. Concerning interpolating operators: most
spectroscopy calculations used only bilocal operators that resemble
noninteracting scattering states, of the form
$D(\vec p_1)D^*(\vec p_2)$ and $D^*(\vec p_1)D^*(\vec p_2)$. However,
Ref.~\cite{Prelovsek:2025vbr} found that it is important to also
include local four-quark operators (i.e.\ with smeared
$cc\bar u\bar d$ at the same point) and that without such operators,
estimates of some energy levels shift significantly. Stump presented
an independent study that confirmed this finding and also showed that
including local operators makes the plateau energy estimates much more
stable with respect to how many bilocal operators are
used~\cite{Stump:2025owq, Stump:2026xvg}.

As discussed in Section~\ref{sec:quantization}, the $DD^*$ partial
wave amplitudes have a left-hand cut close to threshold due to
$u$-channel pion exchange. This was neglected in the earlier
calculations, which used generalized Lüscher quantization conditions
even on this cut where they are invalid. In a reanalysis of the
spectrum from Ref.~\cite{Padmanath:2022cvl} using the plane-wave
approach, Ref.~\cite{Meng:2023bmz} found that the $T_{cc}$ was a
subthreshold resonance rather than a virtual state. Applying the
three-particle approach to the same dataset led to the same
conclusion~\cite{Dawid:2024dgy}. Raposo presented a new calculation
with pion mass 280~MeV and a preliminary analysis using the three-body
approach~\cite{Alharazin:2026lno}, including lattice data for the
relevant two-body subchannels, $I=1$ $DD$ and $I=1/2$ $D\pi$. This
method is particularly suited for the $T_{cc}$ as the pion mass is
decreased toward its physical value, because it is also valid above
the $DD\pi$ threshold. On the other hand, because it treats the $D^*$
as a two-body $D\pi$ bound state, going above the $D^*D^*$ threshold
would require a not-yet-available four-particle formalism. The
preferred quantization condition thus depends on the physics of
interest: at heavier pion masses, two-body methods that include
coupled channels and left-hand cuts can be used to analyze energy
levels above the $D^*D^*$ threshold, whereas at light pion masses the
three-body method is best for the $T_{cc}$.

There have been several additional studies of the $T_{cc}$ and related
doubly charmed systems such as $I=1$ and the singly strange
$DD_s^*$--$D^*D_s$ coupled-channel system, which have not found
additional tetraquark states~\cite{Collins:2024sfi, Meng:2024kkp,
  Shi:2025ogt, Shrimal:2025ues, PitangaLachini:2025pxr,
  Nagatsuka:2025szy}. This includes talks by
Padmanath~\cite{Padmanath_Lat25}, Shrimal~\cite{Shrimal:2026bur}, and
Mohanta~\cite{Mohanta:2026xkz}.

\subsection{Doubly bottom tetraquark}

\begin{figure}
  \centering
  \includegraphics{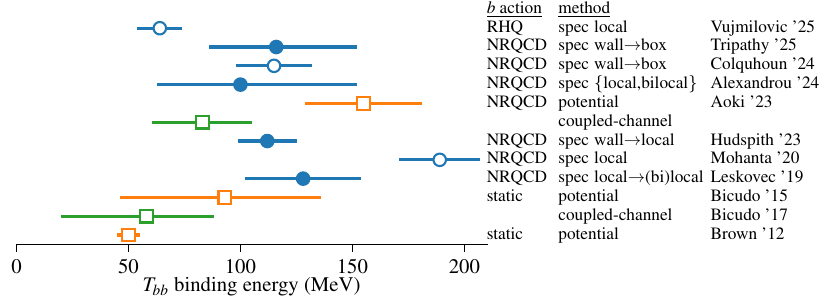}
  \caption{Lattice calculations of the $T_{bb}$ binding energy,
    determined using finite-volume
    spectroscopy~\cite{Leskovec:2019ioa, Mohanta:2020eed,
      Hudspith:2023loy, Alexandrou:2024iwi, Colquhoun:2024jzh,
      Tripathy:2025vao, Vujmilovic:2025czt} (blue circles) and
    potentials via the Born-Oppenheimer~\cite{Brown:2012tm,
      Bicudo:2015vta, Bicudo:2016ooe} and HAL~QCD~\cite{Aoki:2023nzp}
    methods (orange squares for single channel and green squares for
    coupled channel). Open symbols indicate a single lattice spacing
    or a single pion mass was used. Some earlier calculations that
    have been superseded~\cite{Bicudo:2012qt, Francis:2016hui,
      Junnarkar:2018twb} are omitted from this figure.}
  \label{fig:Tbb}
\end{figure}

By replacing the two charm quarks in the $T_{cc}$ with bottom quarks,
we obtain a conjectured state typically called $T_{bb}$ that couples
to $\bar B \bar B^*$. Several independent lattice calculations using
different methods predict a deeply bound state with binding energy of
order 100~MeV, as shown in Fig.~\ref{fig:Tbb} which includes the study
presented by Tripathy~\cite{Tripathy:2026inp}. At the physical pion
mass, the left-hand cut due to pion exchange starts just 1~MeV below
threshold, which makes an analysis of this state using Lüscher's
finite-volume quantization condition
invalid~\cite{Alexandrou:2024iwi}. However, since this state is deeply
bound, we expect exponentially suppressed finite-volume effects.

This doubly bottom tetraquark is a good opportunity for lattice QCD
calculations to make a precise prediction ahead of a possible future
experimental discovery. It is therefore important to carefully examine
possible sources of systematic uncertainty. One reason for concern is
the use of asymmetric correlation functions (typically with wall
sources) in some of these calculations. Ref.~\cite{Alexandrou:2024iwi}
is particularly important for having addressed this by employing a
symmetric correlator matrix with both local and bilocal interpolating
operators. They found that the bilocal operators had no effect on the
estimate of the ground-state energy, which helps to support the
findings of previous calculations. On the other hand, it was recently
shown that using only bilocal operators without local operators yields
a poor estimate of the ground-state energy~\cite{Prelovsek:2025vbr}.

Being a stable hadron makes studying the structure of the $T_{bb}$
much simpler than for exotic resonances. Vujmilovic presented a
calculation of its electromagnetic form
factors~\cite{Vujmilovic:2025czt, Vujmilovic:2026ika}. It was found that
the charge radius of the $T_{bb}$ is smaller than for a $B$ or $B^*$
meson and that the magnetic moment is almost entirely produced by the
$b$ quarks; this was interpreted as favouring a structure in which a
$bb$ $1^+$ diquark is bound in $S$-wave with a $\bar u \bar d$ $0^+$
antidiquark.

\subsection{Other heavy hadrons}

For hadrons containing any combination of light quarks and gluons plus
either two heavy quarks or a heavy quark-antiquark pair, the
Born-Oppenheimer approach is an alternative to the Lüscher method: one
computes potentials for a pair of static (anti)quarks plus the
dynamical light degrees of freedom, then solves a Schrödinger equation
for the heavy quarks. Mohapatra presented a Born-Oppenheimer EFT for
exotic hadrons including the $T_{cc}$ and
$T_{bb}$~\cite{Berwein:2024ztx, Brambilla:2024imu, Brambilla:2025xma,
  Mohapatra_Lat25}, whereas calculations of the potentials were
presented in the talk by Sharma~\cite{Sharma_Lat25} and the poster by
Picão~\cite{Picao_Lat25}.

A few talks covered studies of charmonium. Bali presented a
calculation of $\eta_c$ and $J/\psi$ masses on about 50 ensembles from
CLS, with the usual approximation of no $c\bar c$
annihilation~\cite{Bali_Lat25}. In contrast, Urrea-Niño showed the
effect of removing this approximation on one ensemble with dynamical
charm quarks~\cite{Urrea-Nino:2025afu, UrreaNino_Lat25}.

Stable baryons containing charm and/or bottom quarks, many of which
have not yet been discovered experimentally, were presented in three
talks, all using HISQ ensembles. Shaikh employed a mixed action setup
with NRQCD bottom action, RHQ charm action, and clover light-quark
action~\cite{Shaikh:2026zmk}, whereas Radhakrishnan used HISQ for all
flavours and a very fine lattice~\cite{Radhakrishnan_Lat25}. Dhindsa
presented a precise study of two $\Omega_{ccc}$ baryons using both
overlap and HISQ actions for the valence charm
quarks~\cite{Dhindsa:2024erk, Dhindsa:2025zjk}.

\subsection{Systems with baryons}

The only talk about stable light baryons was by Rosso, who focussed on
the complications arising from $C$-periodic boundary
conditions~\cite{Altherr:2026pux}. Precision calculations of the $\Omega$
mass have played an important role in scale setting, particularly for
the muon $g-2$~\cite{Boccaletti:2024guq, FermilabLattice:2025dui}. In order to
establish strong control over excited-state contributions in these
high-precision calculations, perhaps it may be useful to employ
techniques used for multiparticle spectroscopy.

In the baryon-meson sector, there are an increasing number of studies
of $N\pi$ physics, even to the point of explicitly controlling
intermediate $N\pi$ states in a study of nucleon polarizabilities and
calculating $\gamma^*N\to\pi N$ multipole
amplitudes~\cite{Wang:2023omf, Gao:2025loz}. Paul presented an ongoing
study of $N\to\Delta(1232)$ resonance transition form factors, which
require analytically continuing $N\to N\pi$ transition amplitudes to
the resonance pole~\cite{Paul_Lat25}.

There has been some interest in the $J^P=\frac{1}{2}^-$
$\Lambda(1405)$ resonance, because it has also been suggested to have
a two-pole structure (similar to the $D$ mesons discussed in
Section~\ref{sec:charm_mesons}) with one belonging to the SU(3) octet
and the other to the singlet [now listed as the two-star
$\Lambda(1380)$ in the PDG]. A first study of $\pi\Sigma$--$\bar K N$
coupled channels indeed found both a virtual state and a resonance at
pion mass 200~MeV~\cite{BaryonScatteringBaSc:2023zvt,
  BaryonScatteringBaSc:2023ori}. Mai presented a combined analysis of
those lattice data and experiment using UChPT~\cite{Pittler:2025upn,
  Mai_Lat25b}, whereas an extension of that project to different pion
masses and other negative-parity hyperons was presented by
Alvarado~\cite{Alvarado_Lat25}. Finally, a study with SU(3) symmetry
and pion mass 700~MeV was presented by Sucunza~\cite{Sucunza:2026voz}.

Two studies of baryon-meson scattering at the physical pion mass and
investigations of possible pentaquarks using the HAL QCD method were
reported: $KN$ by Murakami~\cite{Murakami:2025owk, Murakami_Lat25} and
$\bar D N$ by Yamada~\cite{Yamada:2026bfl, Yamada_Lat25}.

Concerning baryon-baryon calculations, I refer to Nicholson's plenary
talk for the current status of $NN$~\cite{Nicholson_Lat25}. Other
talks on $NN$ were given by Perry~\cite{Perry_Lat25},
Chakraborty~\cite{Chakraborty_Lat25}, and
Sugiura~\cite{Sugiura_Lat25}. Channels involving hyperons were
discussed by Laudicina~\cite{BaSc:2026fdy} and
Murase~\cite{Murase_Lat25}.

\section{Outlook}
\label{sec:outlook}

The field of two-hadron spectroscopy using lattice QCD is
maturing. Many hadronic systems of interest have been studied by
multiple independent collaborations. Still, there are some new
methodological developments such as the treatment of left-hand cuts in
finite-volume quantization conditions.

A better understanding of systematic errors is needed. This includes
errors associated with the choice of interpolating operators, plateau
fits or other methods for estimating energy levels, and the modelling
of scattering amplitudes. More investigation of standard lattice
artifacts such as discretization effects would also be valuable: it
was found that the latter can be quite significant in baryon-baryon
systems~\cite{Green:2021qol, Inoue:2024osj, Green:2025rel}. Obtaining
full control over uncertainties is essential for making precise QCD
predictions and making the connection with experiment.

For many hadrons, an obstacle to decreasing the pion mass towards its
physical value is the opening of new channels with three or more
hadrons. The necessary three-hadron spectroscopy is a relatively new
field that has made significant progress in recent years but with much
pioneering work still to come.

\acknowledgments

I thank everyone who communicated with me about their results:
J.~Baeza~Ballesteros, D.~Chakraborty, X.~Feng, M.~Mai, N.~Mathur,
K.~Murakami, M.~Padmanath, R.~Perry, A.~Portelli, S.~Prelovšek,
J.~T.~Tsang, F.~Romero-López, M.~Salg, T.~Scirpa, J.~A.~Urrea-Niño,
and I.~Vujmilovic.

The correlation functions used to produce Fig.~\ref{fig:Tcc_plateaus}
were computed using resources provided by the Gauss Centre for
Supercomputing e.V.\ (\url{www.gauss-centre.eu}) on
JUWELS~\cite{juwels} at Jülich Supercomputing Centre. This work used
the software packages QDP++~\cite{Edwards:2004sx},
PRIMME~\cite{PRIMME}, QUDA~\cite{Clark:2009wm, Babich:2011np,
  Clark:2016rdz, Clark:2025cuz}, NumPy~\cite{numpy},
SciPy~\cite{scipy}, CuPy~\cite{cupy}, and
Matplotlib~\cite{Hunter:2007}.

\bibliographystyle{JHEP}
\bibliography{spectroscopy}

\end{document}